# Can Musical Emotion Be Quantified With Neural Jitter Or Shimmer?
# A Novel EEG Based Study With Hindustani Classical Music

Sayan Nag, Sayan Biswas, Sourya Sengupta
Department of Electrical Engineering
Jadavpur University
Kolkata, India

Shankha Sanyal, Archi Banerjee, Ranjan Sengupta, Dipak Ghosh
Sir C.V. Raman Centre for Physics and Music
Jadavpur University
Kolkata, India

*Abstract*—The term jitter and shimmer has long been used in the domain of speech and acoustic signal analysis as a parameter for speaker identification and other prosodic features. In this study, we look forward to use the same parameters in neural domain to identify and categorize emotional cues in different musical clips. For this, we chose two *ragas* of Hindustani music which are conventionally known to portray contrast emotions and EEG study was conducted on 5 participants who were made to listen to 3 min clip of these two *ragas* with sufficient resting period in between. The neural jitter and shimmer components were evaluated for each experimental condition. The results reveal interesting information regarding domain specific arousal of human brain in response to musical stimuli and also regarding trait characteristics of an individual. This novel study can have far reaching conclusions when it comes to modeling of emotional appraisal. The results and implications are discussed in detail.

## I. INTRODUCTION

Music is nothing but a dynamic signal with a proper and rhythmic mixture of different sounds and silence. Music or audio signals give rise to different types of emotions in human brain[1]. Thus, in cognitive neuroscience brain response to different musical stimuli is becoming an interesting field of research day by day. Different types of music have different impacts on human brain like; effect of Rock music will not be the same with Hindustani classical music.

In North Indian Classical Music, *raga* [2] forms the basic structure over which individual improvisations is performed by an artist based on his/her creativity. The *raga* is a sequence of musical notes and the play of sound which delights the hearts of people. The word *Raga* is derived from the Sanskrit word "Ranj" which literally means to delight or please and gratify. Although there are a number of definitions attributed to a Raga, it is basically a tonal multifarious module.

In India, music (*geet*) has been a subject of aesthetic and intellectual discourse since the times of *Vedas (samaveda)*. *Rasa* was examined critically as an essential part of the theory of art by Bharata in Natya Sastra, (200 century BC). The *rasa* is considered as a state of enhanced emotional perception produced by the presence of musical energy. It is perceived as a sentiment, which could be described as an aesthetic experience. Although unique, one can distinguish several flavors according to the emotion that colors it [11]. Several emotional flavors are listed, namely erotic love (*sringara*), pathetic (*karuna*), devotional (*bhakti*), comic (*hasya*), horrific (*bhayanaka*), repugnant (*bibhatsa*), heroic (*vira*), fantastic, furious (*roudra*), peaceful (*shanta*) [10].

To interpret and analyze human response of a particular emotional musical stimuli Electroencephalogram (EEG) is a useful tool  A number of bio-sensors are attached with head through some electrical probes to record the voltage variation over time. This is very much useful to decode the function of complex human brain structure. EEG signals can be divided into some major categories according to frequency range (i) delta (δ)  0-4Hz, (ii) theta (θ)  4–8 Hz, (iii) alpha (α) 8-13Hz and (iv) beta (β) 13-30 Hz. In the previous works it has been found that different types of emotions have different impact of different bands of EEG signals[3].

In this work, we are going to present a novel study regarding the variation of neural jitter and shimmer of EEG responses corresponding to two musical stimuli (Raga Chayanat, Raga Darbari) of contrast emotions [11]. Previously no work has been done as per our knowledge, which deals with the application of jitter and shimmer techniques in bio-signals to assess emotional cues. Jitter conventionally refers to the variability of fundamental frequency while shimmer refers to the variability in the peak to peak amplitude [12]. Applying the same concept to neural EEG domain, we have termed this as neural-jitter and neural-shimmer.

The rest of the paper has been organized as follows: Section II depicts the details about the Theory of jitter and shimmer for normal signals. Materials and data used for this work is described in section III. Approach of applying neural jitter and neural shimmer is presented in section IV. Section V reveals the results obtain during the analysis.

## II. THEORY

### A. Jitter and Shimmer

The most important acoustic parameters used to measure the quality of voice are jitter and shimmer[4]. These parameters are being calculated to measure the perturbation index in a signal. Here, we calculated the jitter and shimmer parameters for Electroencephalogram signals and compared them for various EEG signals.

### B. Measurement of Shimmer Values:

*Shimmer* is the variation of peak-to-peak amplitude. It is defined as the mean absolute difference between amplitudes of successive periods divided by the mean value of the amplitudes. Mathematically, given as

$$\frac{\frac{1}{N-1}\sum_{i=1}^{N-1}|A_i - A_{i+1}|}{\frac{1}{N}\sum_{i=1}^{N}A_i}$$

Where $N$ is the number of periods and $A_i$ is the amplitude.

### C. Measurement of jitter values:

Jitter is the perturbation of fundamental frequency. It can also be defined as the deviation from true periodicity of an apparently periodic signal, given as the mean absolute difference between successive periods divided by the mean value of the periods. Mathematically, given as

$$\frac{\frac{1}{N-1}\sum_{i=1}^{N-1}|T_i - T_{i+1}|}{\frac{1}{N}\sum_{i=1}^{N}T_i}$$

Where $N$ is the number of periods and $T_i$ is the duration (in seconds) of the $i^{th}$ period.

## III. MATERIALS AND METHODOLOGY

### A. Subjects Summary:

Five musically untrained male young adults voluntarily participated in this study. The average age was 23 (SD = 1.5 years) years and average body weight was 70kg. Informed consent was obtained from each subject according to the guidelines of the Ethical Committee of Jadavpur University. All experiments were performed at the Sir C.V. Raman Centre for Physics and Music, Jadavpur University, Kolkata. The experiment was conducted in the afternoon with a normal diet in a normally conditioned room sitting on a comfortable chair and performed as per the guidelines of the Institutional Ethics Committee of SSN College for Human volunteer research. This is a pilot study as part of an ongoing research project at the Centre and the number of EEG respondents will be increased gradually.

### B. Experimental Details:

EEG recording was done with two *ragas* – Chayant and Darbari Kanada of Hindustani music for the 5 subjects. From the complete playing of the ragas, segments of about 2 minutes were cut out for analysis of each raga. The emotional part of each clip was identified with the help of experienced musicians. Listening test was conducted beforehand with 30 participants to standardise the emotional content of each musical clip. 75% of the participants found Chayanat to be joyful, 80% found Darbari to convey pathos emotion. These findings were corroborated with the brain response data.

Variations in the timbre were avoided by making the same artist play the two ragas with the same sitar. Amplitude normalization was done for both the signals therefore loudness cues were not present. Each of these sound signals was digitized at the sample rate of 44.1 KHZ, 16 bit resolution and in a mono channel. A sound system (Logitech R _ Z-4 speakers) with high S/N ratio was used in the measurement room for giving music input to the subjects.

### C. Experimental Protocol:

EEG was done to record the brain-electrical response of 5 subjects. Each subject was prepared with an EEG recording cap with 19 electrodes (Ag/AgCl sintered ring electrodes) placed in the international 10/20 system. Fig. 1 depicts the positions of the electrodes.

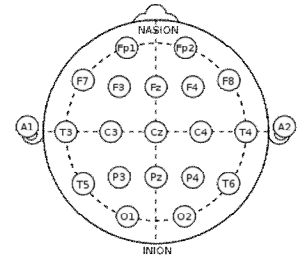

Fig. 1 The position of electrodes according to the 10-20 international system

Impedances were checked below 5 kOhms. The EEG recording system (Recorders and Medicare Systems) was operated at 256 samples/s recording on customized software of RMS. The data was band-pass-filtered between 0.5 and 35Hz to remove DC drifts and suppress the 50Hz power line interference. Each subject was seated comfortably in a relaxed condition in a chair in a shielded measurement cabin. They were also asked to close their eyes. On the first day after initialization, a 14 min recording period was started, and the following protocol was followed:
1. 2 min No Music,
2. 2 min With Drone

3. 2 min With Music 1 ( Chhayanat ),
4. 4. 2 min No Music
5. 2 min With Music 2 (Darbari Kannada)
6. 2 min No Music

The drone signal has been used as a baseline over which the emotional arousal corresponding to other musical clips have been taken. The tanpura drone creates a repetitive buzzing sound which helps to create an atmosphere without evoking any specific emotion [6].

*D. Methodology:*

In order to eliminate all frequencies outside the range of interest, data was band pass filtered with a 0.5-35 Hz FIR filter. The amplitude envelope of the alpha (8-13 Hz), theta (4-7 Hz) and gamma (30-50 Hz) frequency ranges were obtained using wavelet transform technique The amplitude envelope of the different frequency rhythms were obtained for 'before music', 'with music' as well as 'without music' conditions for for a total of 19 electrodes which are Fz, Cz, Pz, Fp1, Fp2, F3, F4, C3, C4, P3, P4, O1, O2, F7, F8, T3, T4, T6.

## IV. APPROACH

EEG signals (Fig.1) are obtained in the temporal domain showing the variation of Extracellular Field Potentials [5] with respect to time. EEG signals has played a crucial role in the domain of music cognition [6] [7] [8]. Digital Signal processing methods allows converting between temporal domains to frequency domain. The transform is called Fourier Transform [9] which generates Fourier coefficient for each frequency present in the temporal signal. The plot of the frequency domain, which shows the Fourier coefficients, is called frequency spectrum, (Fig.2). The work is aimed to look into proposing a novel technique to calculate traditional Jitter Shimmer in neural signals. The EEG signals were transformed into frequency domain double sided frequency spectrum (Fig. 3). The Fourier coefficients can be both real and complex. The plot and analysis was made with the absolute values of the coefficients. Single sided spectrum (Fig.4) was used for analysis instead of double sided. An attempt to analyze the single sided frequency spectrum is made by considering only top 20% frequency coefficients. Threshold is 20% of ($A_{max}$) where is $A_{max}$ the largest Fourier coefficients (Fig.5). The Fourier coefficients above $A_{max}/5$ continue to possess its values and below $A_{max}/5$ becomes zero (Fig.6). This transformation is expected to be good as values less than $A_{max}/5$ are literally less so losing those coefficient is expected not to cause much loss of information. A Frequency thresholding (*FT*) matrix is computed storing values of time periods ($2*pi*\omega$), $\omega$ are the frequencies with non-zero Fourier coefficients in thresholding curve) and the corresponding Fourier coefficients. Fig.7 shows the plot with non-zero coefficients only. The time elements in *FT* matrix serve as $T_i$ in calculating the jitter and the corresponding coefficient elements in *FT* matrix serve as $A_i$ in the shimmer calculation.

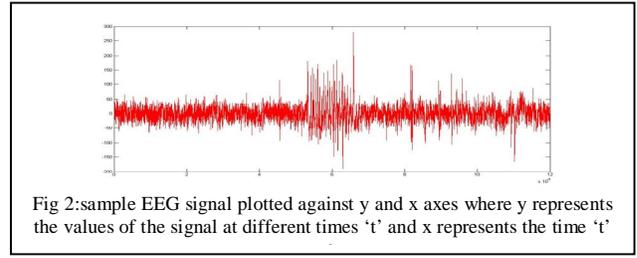

Fig 2:sample EEG signal plotted against y and x axes where y represents the values of the signal at different times 't' and x represents the time 't'

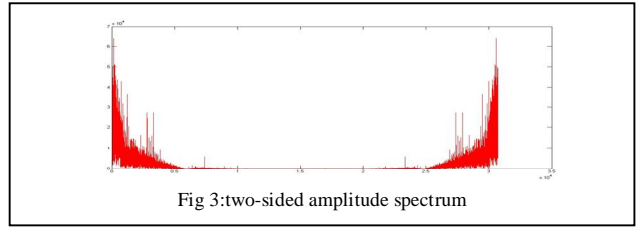

Fig 3:two-sided amplitude spectrum

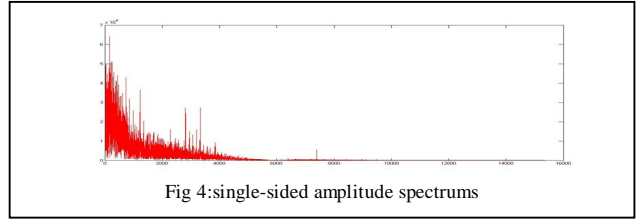

Fig 4:single-sided amplitude spectrums

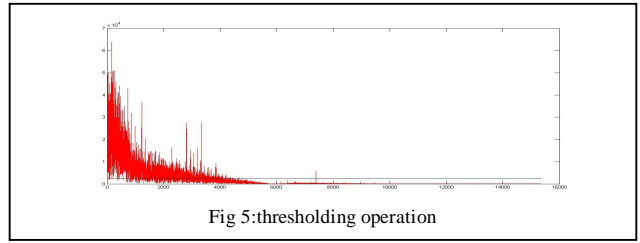

Fig 5:thresholding operation

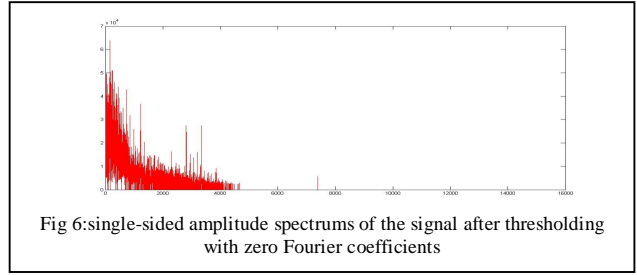

Fig 6:single-sided amplitude spectrums of the signal after thresholding with zero Fourier coefficients

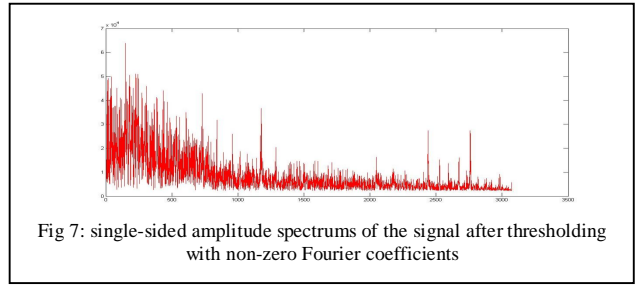

Fig 7: single-sided amplitude spectrums of the signal after thresholding with non-zero Fourier coefficients

## V. RESULTS

The jitter and shimmer values were calculated for each of the experimental conditions as illustrated in the Experimental Protocol Section. The neural jitter computed showed very little variance acrosss the six different experimental conditions; the variance occuring mostly after the fourth places of decimal. So we can safely assume that there is negligible or almost no change in the neural jitter values under the effect of various stimuli, and therefore it can be considered as a source characterisitic. The neural shimmer values, however showed significant changes under the effect of different emotional stimuli and could prove to be a robust parameter for categorization and classification of perceived musical emotions. In the following figures (**Fig. 8-13**) the variation of neural shimmer values have been plotted for different frequency ranges under the effect of a particular stimuli and also what happens after the removal of that stimuli. The after stimulus part have been incorporated to have a look at the retentive capacities of the different lobes of brain when the music stimulus have been removed.

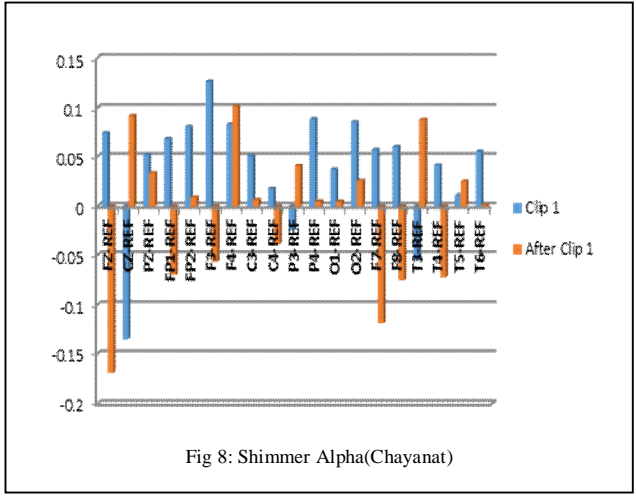

Fig 8: Shimmer Alpha(Chayanat)

In case of alpha frequency range, we find that the shimmer values increase for most of the electrodes under the effect of *raga Chayanat*. The increase is most prominent in the frontal electrodes with the highest being noted in the left frontal F3 electrode, indicating the greater involvement of left frontal lobe in the processing of positive emotions. The neural shimmer also increases quite consistently in the two fronto-parietal electrodes (FP1 and FP2) as well as in the right parietal (P4) and occipital (O2) electrodes. The central midline electrode (Cz) registers a significant dip in neural alpha shimmer along with left temporal (T3) one. Another interesting observation that comes from the plot is that except for a few electrodes, the aroused level of neural shimmer does not change even after the removal of stimuli, but is retained for some time. Except for few electrodes like Fz, Cz and F7, F8 where there is significant dip in the alpha neural shimmer values after the removal of stimulus. Thus, we can say that a *raga* of positive emotion increases the neural simmer of alpha range mostly in a positive manner and that is retained for quite some time even after the removal of the stimuli.

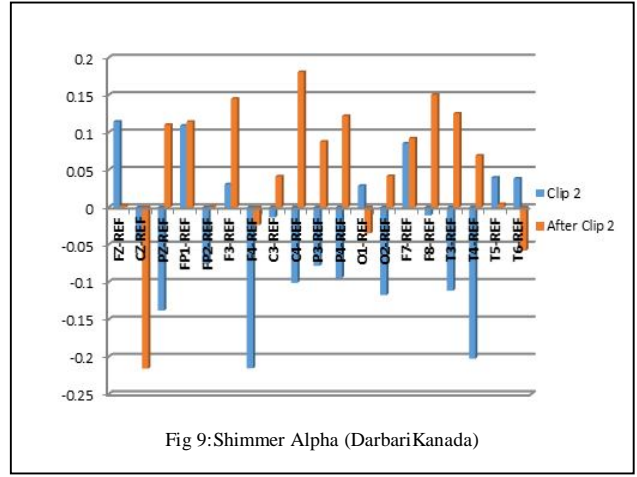

Fig 9: Shimmer Alpha (DarbariKanada)

In case of *raga Darbari,* conventionally associated with sad emotion, we find the effect in neural shimmer is exactly in contrast with the first case. Here, we find that the neural shimmer values are decreasing significantly under the effect of *raga Darbari* in most of the electrodes, with the dip being most significant in case of right frontal (F4) and temporal (T4) electrode. This may be an indication in the direction of negative emotions being processed in the right hemisphere of brain. The right central (C4), parietal (P4) and occipital (O2) electrodes also display consistent fall in the neural shimmer values. In this case, we also see that the retention is not so pronounced as in the previous case and the neural shimmer values increase consistently after the removal of stimuli.

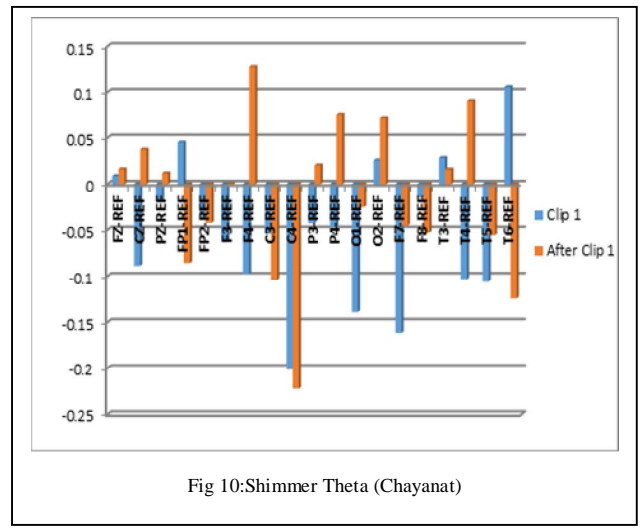

Fig 10: Shimmer Theta (Chayanat)

In case of theta frequency range, we find that the shimmer values decreases for most of the electrodes under the effect of *raga Chayanat*. The decrease is most prominent in the right central (C4) electrode, followed by the O1, F7, F4, T4 and Cz electrodes. The right temporal, i.e., the T6 electrode shows a significant increase in neural shimmer value. After removal of stimuli the dip in the neural shimmer values is retained for few electrodes, especially which is significant for the right Central (C4) electrode which also showed the most drop while the clip was on. Also, the right frontal electrode F4 shows a significant rise in the neural shimmer value after the removal of stimuli. Thus, we can say that a *raga* of positive emotion decreases the neural shimmer of theta range.

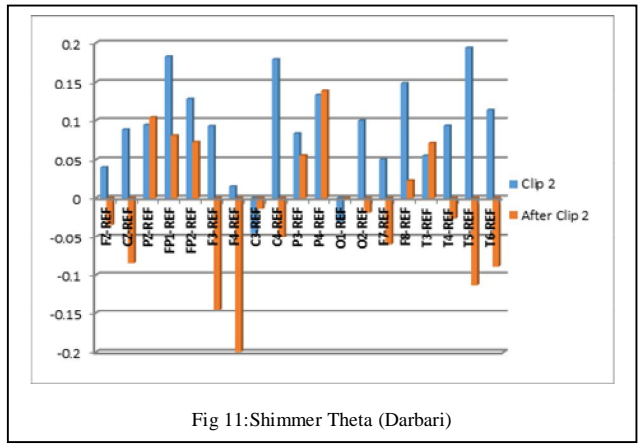

Fig 11:Shimmer Theta (Darbari)

In case of theta frequency range, we find that the shimmer values increase for most of the electrodes under the effect of *raga Darbari* which is just in contrast with that of *raga Chayanat*. The increase is prominent in the right central (C4), left fronto-parietal (Fp1) and left temporal (T5) electrodes, followed by FP2, P4, F8,T6 and O2 electrodes. The left central (C3) electrode shows a decrease in neural shimmer value along with left Occipital (O1) electrode. After the removal of stimuli the aroused level of the neural shimmer values is retained for few electrodes, while there is a significant dip in the neural shimmer value for right frontal (F4) electrode which in case of *raga Chayanat* showed an increased value of neural shimmer.

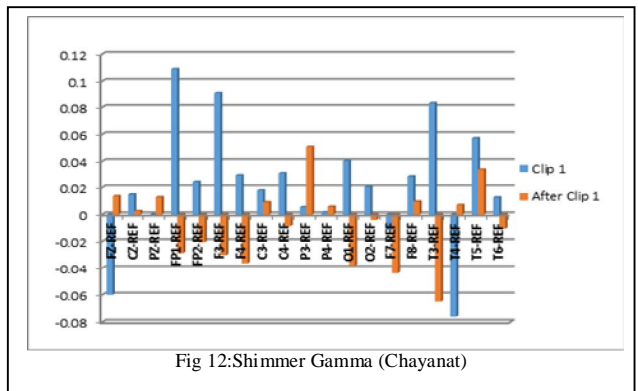

Fig 12:Shimmer Gamma (Chayanat)

In case of gamma frequency range, we find the shimmer values increase for most of the electrodes under the effect of *raga Chayanat*. The increase is most prominent in most of the electrodes with the highest being noted in the left fronto-parietal Fp1 electrode, which implies that fronto-parietal lobe plays an important role in the processing of positive emotions. The neural shimmer also increases quite consistently in the two fronto-parietal electrodes (FP2 and FP3) as well as in the left temporal (T3) and occipital (O1) electrodes. The frontal midline electrode (Fz) registers a significant dip in neural gamma shimmer along with right temporal (T4) one. Another interesting observation that comes from the plot is that except for a few electrodes, the aroused level of neural shimmer changes after the removal of stimuli except for few electrodes like Fz, Pz and T5, P3 where there is significant rise in the gamma neural shimmer values after the removal of stimulus. Thus, we can say that a *raga* of positive emotion increases the neural shimmer of gamma range mostly in a positive manner.

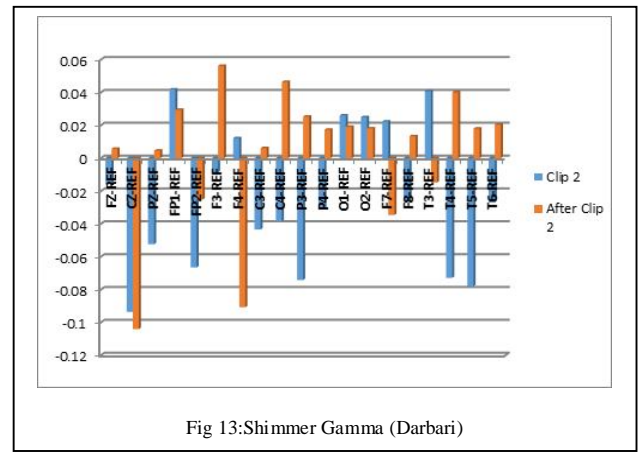

Fig 13:Shimmer Gamma (Darbari)

In case of *raga Darbari* for gamma frequency range, we find the effect in gamma neural shimmer is exactly the opposite with what we have found in case of *raga Chayanat*. Here, we find that the neural shimmer values are decreasing significantly under the effect of *raga Darbari* in most of the electrodes, with the dip being most significant in case of central midline (Cz), left temporal (T5) and right temporal (T6) electrodes. The left and right central (C3, C4) and parietal (P3, P4) electrodes also display consistent fall in the neural shimmer values. The left and right occipital (O1 and O2) shows an increase in the shimmer values along with Fp1 and T3. In this case, we also see that the retention is not pronounced after the removal of stimuli but Cz electrode shows the maximum drop in the shimmer value both before and after the removal of clip.

Jitter is conventionally defined as the cycle-to-cycle variation of fundamental frequency, that is, the average absolute difference between consecutive periods divided by the average value of the individual periods, expressed as in equation 2.When applied to the neural domain, jitter may be considered as a measure of perturbation suffered by the fundamental EEG frequency; since this perturbation is essentially constant for a particular person, we could not find significant variations in the jitter parameter under the influence of a variety of stimuli. But, the analysis may be carried out differently using other parametric measures to evaluate neural jitter which may have definite far reaching conclusions in the modeling of emotions.

## CONCLUSION

In this work, we propose two novel parameters – neural jitter and shimmer whose variations has been used as a parameter to quantify and categorize emotional arousal using Hindustani classical music as a stimuli. The study yields the following interesting conclusions:

1. The neural jitter is a subjective parameter which is very much dependent on the state of consciousness of a particular person and thus remains mostly unaffected by any type of emotional music stimuli. In the domain music signal analysis, jitter has largely been used as a parameter which defines the timbral characteristic of a musical instrument. In the same way, we propose to use neural jitter as a parameter which defines the state of consciousness of a human being.

2. The neural shimmer has been calculated for different frequency bands of EEG, reveals interesting data regarding the arousal and retention of different emotions in human brain. In the alpha frequency range, elevated levels of neural shimmer are representation of positive emotion while diminished levels are marker for negative emotion. The retention is higher in case of positive emotion as compared to negative.

3. In case of theta frequency range, just the opposite response is found in case of arousal by positive and negative emotional music. The central lobe seems to be the most affected by positive stimuli while the temporal and fronto-parietal electrodes are mostly affected in case of negative stimuli. Again, the retention based effects are more prominent in case of the $1^{st}$ clip as compared to the $2^{nd}$.

4. The gamma frequency range reveals almost the same result as the alpha range, with elevated and diminished levels are markers of positive and negative emotional stimuli respectively. The left fronto-parietal, frontal and temporal lobe plays significant part in the processing of positive emotion, while the central, parietal and temporal lobes are strongly associated with negative emotions.

In this way, we propose a novel algorithm which can be utilized for quantification and categorization of emotional arousal in respect to Hindustani classical music. More rigorous works being carried out in this domain include the variation of neural jitter or shimmer values within the span of particular emotional stimuli- i.e. we envisage to characterize the fluctuation of fluctuations. Analysis with a variety of other musical stimuli and greater number of participants would yield a more statistically significant and conclusive result and conclusion. This is a pilot study in that direction.

## ACKNOWLEDGMENT

One of the authors, AB acknowledges the Department of Science and Technology (DST), Govt. of India for providing (A.20020/11/97-IFD) the DST Inspire Fellowship to pursue this research work. Another author, SS acknowledges the West Bengal State Council of Science and Technology (WBSCST), Govt. of West Bengal for providing the S.N. Bose Research Fellowship Award to pursue this research (193/WBSCST/F/0520/14).

All the authors acknowledge Department of Science and Technology, Govt. of West Bengal for providing the RMS EEG equipment as part of R&D Project (3/2014).